\documentstyle[twoside,fleqn,espcrc2,epsbox]{article}
\newcommand{\AmS}{{\protect\the\textfont2
  A\kern-.1667em\lower.5ex\hbox{M}\kern-.125emS}}

\title{Phase structure of ${\rm CP}^{N-1}$ model with topological  
term \thanks{Talk presented  by H. Yoneyama. SAGA-HE-138, YAMAGATA-HEP-98-17}}

\author{Masahiro Imachi,\address{Department of Physics, Yamagata 
University,
 Yamagata , Japan}%        
                Shouhei Kanou\address{Hitachi Tohoku Software Ltd., 
                Sendai, Japan}
        and 
        Hiroshi Yoneyama\address{Department of Physics, Saga University, Saga, Japan}
        }
       
\begin{document}

\begin{abstract}
${\rm CP}^{N-1}$ model with topological term is numerically studied.  
The topological charge distribution $P(Q)$  is calculated and then 
 transformed to the partition function Z($\theta$) as a function of $\theta$
  parameter.
In the strong coupling region, $P(Q)$ shows a gaussian behavior, which 
indicates a   first order phase transition 
at $\theta =\pi$. In the weak coupling region, 
 $P(Q)$ deviates from gaussian. 
A bending behavior of resulting $F(\theta)$ at $\theta \neq \pi$, which 
might  be a signal of a first
 order phase  transition, could  be misled  by large errors coming 
 from the fourier transform of  $P(Q)$. Results are shown mainly  for
  ${\rm CP}^{3}$ case.
\end{abstract}

\maketitle

\section{Introduction}

       It is well known that ${\rm CP}^{N-1}$ model shares with QCD many
    dynamical aspects such as asymptotic freedom, confinement,
   dynamical mass generation and  non-trivial topology.   We are 
   concerned with the topological aspects of the model and study the 
   nature of the $\theta$ vacuum.  It is deeply associated with the
   non-perturbative nature of the strong interaction. 
   The strong $CP$ problem is one of the issues to be clarified 
   nonperturbatively.   
  It is also expected that a new parameter introduced into the 
  theory,   in general, would induce a rich phase structure. 
  Actually, in 1982, Cardy took,  as a toy model,  the Z(N) gauge model 
  and by making a renormalization group argument showed that very rich phase 
  structures emerged \cite{Car}.
In this talk we are concerned with the dynamics of  the  $\theta$ 
  vacuum of the    ${\rm CP}^{N-1}$ model  with the topological term, 
  and report on the  results mainly of ${\rm CP}^{3}$.  
  ${\rm CP}^{3}$ model is semiclassically   free from    dislocation. \par
     From the numerical point of view,  the topological term introduces
 complex Boltzmann weight in the euclidean space time.   It prevents one from 
applying straightforwardly the standard algorithm of the Monte Carlo
 simulations.   This problem can be circumvented by Fourier-transforming 
 the topological charge distribution $P(Q)$ \cite{BDSL,W89,OS,BPW,HITY,PS}.
   It is then necessary to 
 calculate P(Q) in a high precision because Fourier transformation 
 will generate large error propagation.
 In order to achieve a  high precision, we employ multi-histogram 
 method as well as reweighting method.\par
\section{Formulation} 
    The  lattice action of  the ${\rm CP}^{N-1}$ model is given by
\begin{equation}
 S=  -\beta N \sum_{n,\mu}  {\overline z_{n+\mu}} U_{n,\mu} z_n, + 
 {\rm c.c.}	
 	\label{eqn:e1}
 	\end{equation}
 where $z_{n}$ is CP$^{N-1}$ variable at site $n$ and  $U_{n,\mu}$ is
  U(1)  variable sitting on a link $n,\mu$.  $\beta$ is the
 coupling constant.  This action is known to be superior with respect 
 to  scaling behavior  to  the standard action \cite{CRV},  
 which  is quartic in $z_{n}$ variables.
   We use metropolis 
 algorithm combined with the overrelaxation,  which is applicable 
 to  arbitrary $N$.   
 The $\theta$ term is added to  eq.(\ref{eqn:e1})~.
  \begin{equation}
 S_\theta= S -  i \theta \hat{Q}
 \end{equation} 
 where the topological charge is computed according to the 
 definition 
\begin{eqnarray}
\hat{Q}={1 \over 4\pi } \sum_{n,\mu,\nu} \epsilon_{\mu \nu}
 ( \theta_{n,\mu} +  \theta_{n+\mu,\nu} -
\theta_{n+\nu,\mu} - \theta_{n,\nu} ),
 \end{eqnarray}
%
%where $\theta_{n,\mu}= {\rm arg} \{ {\overline z_n}z_{n+\mu} \}$.\par
 where $\theta_{n,\mu}$ is the phase of $U_{n,\mu} $.\par
  In order to avoid  complex Boltzmann weights, we adopt the 
  algorithm by which the partition function is given by the Fourier 
  transform of the topological  charge distribution  $P(Q)$
  \begin{equation}
      	Z(\theta)=\sum_{Q}e^{i \theta Q} P(Q)
      	\label{z}
      \end{equation}   
 The distribution  $P(Q)$ is calculated by the real Boltzmann weight 
 \begin{equation}
    	P(Q)=\int [dz d{\overline z}]^{Q}e^{-S}/\int [dz d{\overline z}]e^{-S}
    	\label{pq}
    \end{equation}
    where $ [dz d{\overline z}]^{Q}$ is the constrained   measure in which the 
   value of the  topological charge  is restricted to  $Q$, 
   and $P(Q)$ 
   is normalized such that   $\sum_{Q} P(Q)=1$.
 %
% \subsection{set method and  trial function }  
       Error propagation in the fourier transform (\ref{z})~,    
    in general, causes large errors to $Z(\theta)$, free energy 
    and its derivatives.  In order to acquire reasonable results 
    within  tolerable errors, 
    we   then have to  calculate $P(Q)$  in a 
    very high precision. For that purpose, a  combined use of  the 
    multi-histogram method (set method) and  reweighting method (trial function
     method) is made.  Since $P(Q)$  is an even function of $Q$,  we 
     divide the range of positive (and 0) values of 
     $Q$ into sets $S_{i}, i=1, 2, 3,\ldots$.  Each of the sets $S_{i}$
      consists of 4 bins $Q=3i-3, 3i-2, 3i-1, 3i$ so that adjacent 
      sets  overlap at the edge bins of each set,  $Q=3k (k=1,2,3,\ldots)$.     
For negative values of $Q$, we use $P(Q)=P(-Q)$.
\bigskip
\section{Results} 
\subsection{Strong coupling region}
  Topological charge distribution $P(Q)$  is calculated for various 
  volumes $V$ and coupling constants.  Typical behavior of $P(Q)$ 
  in the strong coupling region is plotted as a function of $Q^{2}$
   in Fig.1 for a fixed  $\beta$.  
    Solid lines are gaussian fittings to
   the data, and it is clearly seen  that the fittings work very well for 
   all $V$. 
%
%\psbox[height=5.5cm,width=6cm]{Fig1.eps}
\begin{figure}[htb]
\vspace{9pt}
\framebox{\psbox[height=4.0cm,width=7cm]{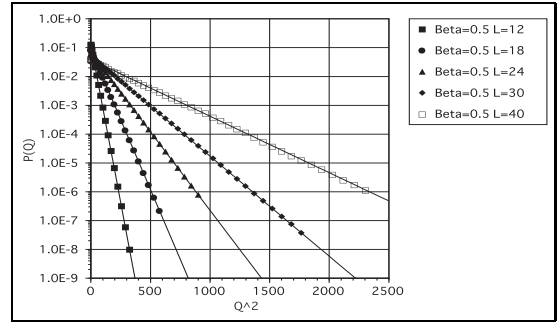}}
\caption{$P(Q)$ at $\beta=0.5$ for  ${\rm CP}^{3}$.}
\label{fig:Fig.1}
\end{figure}
   The coefficient $a_{2}$, read off from  
   the fit  $P(Q)=A\exp\left(-a_{2}Q^{2}
   \right)$,  shows
    that $a_{2} \propto 1/V \equiv C/V$ ($C$ is a constant), which turns out to be a finite size scaling 
    law of  the first order     phase transition at $\theta=\pi$. 
     For that,  the  partition function 
     is now  expressed as the third elliptic
     function $\vartheta_3$, 
     	$Z(\theta)\propto \vartheta_3(\nu,\tau)$, 
		where $\exp(i2\pi\nu) \equiv e^{i\theta}$ and $\exp(i\pi\tau) \equiv \exp(-C(\beta)/V)$.
	Infinite product 
	$$\vartheta_3(\nu,\tau)\propto
	\prod_{n=1}^{\infty}
	\left(1+p^{2n-1} \zeta\right)\left(1+\frac{p^{2n-1}}{\zeta}\right)$$
    yield zeroes at $\zeta=- p^{-\left(2n-1\right)}, - p^{2n-1}$ or 
     $\theta=\pi\pm i \frac{(2n-1)C}{V}$ ( $n=1,2,3,\ldots$ ), where  
    $ \zeta \equiv \exp(i2\pi\nu)$ and $p\equiv \exp(i\pi\tau)$. 
    Accumulation of zeroes  according to the  finite size scaling 
    law $1/V$ indicates a first order phase transition at  
    $\theta=\pi$ \cite{IPZ}. \par 
    Resulting free energy density $	F(\theta)=\ln Z(\theta)/V$ 
 develops a cusp  at $\theta=\pi$  as $V$ increases because of the periodicity 
 $F(\theta+2\pi)=F(\theta)$. Its  behavior 
 agrees  with the result of the strong coupling expansion by 
 Seiberg\cite{Sei}.

\subsection{Weak coupling region}   
  In the weak coupling region, $P(Q)$ deviates from the gaussian.
We tried various fits to $\ln P(Q)$ such as Polynomial fits, or 
adding $Q^{1/2}, Q^{3/2}$ etc.  to them  in order to make the fourier transform 
efficient.  However, none of our trials  was successful in the sense 
that resulting errors of   $F(\theta)$ is so large that meaningful 
results could not be obtained.   Although there may be some other better fits, we decide
 to use the   data itself for the Fourier transform.  \par   
Typical behavior of the resulting free energy is shown in Fig. 2.  $F(\theta)$ 
develops  a sharp bend at $\theta_{b}\neq\pi$.  
At the conference, we reported on a possibility of first 
 order phase transition at $\theta\neq\pi$. 
Thereafter, however, we came to realize an insufficiency  of our 
 way of analysis. So the conclusions we drew in the talk  might have 
 been somewhat misleading. 
 The partition function receives large errors from $P(Q)$, 
 particularly from $P(Q)$ in the first set $Q=0,1,2$ and $3$,  because 
 of rapidly decreasing function.
 $Z(\theta)$ becomes then  meaningful only for the values of 
 $\theta$  
 such that $Z(\theta) > |\delta Z|$, where $\delta Z$ is a 
 fluctuation coming from the fourier transform of $P(Q)+\delta P(Q)$.
By setting  $\theta \equiv \theta_{fl}$ such that  $Z(\theta_{fl}) \approx  
|\delta Z|$, data seem to  indicate $\theta_{b}\approx \theta_{fl} $. 
 Since $\theta_{fl}$ is associated with 
  the errors of $P(Q)$, its location depends on the number of 
  statistics.  Our preliminary analysis seems to support this  
  feature. \par
\begin{figure}[htb]
\vspace{9pt}
\framebox{\psbox[height=4.0cm,width=7cm]{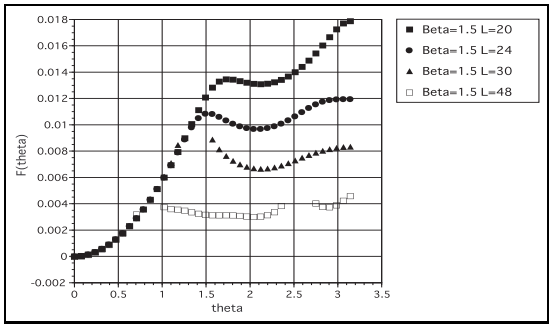}}
\caption{ $F(\theta)$ at $\beta$=1.5 for  ${\rm CP}^{3}$. Errors are not included. }
\label{fig:Fig.2}
\end{figure}
\section{Comments}
We need to be more careful with the weak coupling 
region. Bending behavior of $F(\theta)$ at $\theta\neq\pi$ in the weak coupling region 
might not lead straightforwardly to the conclusion that  first order phase 
transition occurs  at $\theta\neq\pi$ \cite{PS}.  
  Our results of volume dependence of $\theta_{b}$ for various  $\beta$ 
  is    quite similar to that obtained by Schierholz.  
  In his  papers \cite{OS}  Schierholz  concluded that $\theta_{c}$  linearly 
   decreases in 
  $V$, and that its extrapolation to $V\rightarrow\infty$   may  lead 
  $\theta_{c}\rightarrow 0$ as a possible solution to the strong $CP$
  problem.  Our data are, however, too noisy to draw such a 
  conclusion, and the detail will be reported in \cite{IY}.  \par
   In addition to  ${\rm CP}^{3}$ model, we  reported about the results of 
    ${\rm CP}^{1}$ and  ${\rm CP}^{2}$ models with  
the standard actions,  quartic in $z$ variables.
In case of  ${\rm CP}^{1}$, fixed point action is  also used for computations. 
Compared to the standard action, fixed point action  shows that strong coupling behaviors are seen 
only in the  restricted region very close to  $\beta=0$.  
As far as the behavior of $F(\theta$) is concerned, these models show 
similar behaviors in both of the strong and weak coupling regions.
\par
The authors thank R. Burkhalter for useful discussions.
 This work is supported in part by the Grant-in-Aid for Scientific 
Research from the Japanese Ministry of Education, Science and 
Culture (No. 10640276). 

\end{document}